\documentclass[
aps,%
12pt,%
final,%
notitlepage,%
oneside,%
onecolumn,%
nobibnotes,%
nofootinbib,%
superscriptaddress,%
centertags]%
{revtex4}

\newcommand{\bi}{\bibitem}
\newcommand{\be}{\begin{eqnarray}}
\newcommand{\ee}{\end{eqnarray}}
\newcommand{\rar}{\rightarrow}

\begin{document}
%\selectlanguage{english}

\title{Cosmic antigravity}
\author{\firstname{A.~D.}~\surname{Dolgov}}
\email{dolgov@itep.ru}
\affiliation{Novosibirsk State University, Novosibirsk, Rusia}
\affiliation{ A.I.~Alikhanov Institute of Theoretical and Experimental Physics, Moscow}
\affiliation{University of Ferrara and INFN, Ferrara, Italy
}

\begin {abstract}
Possibility of gravitational repulsion  in General Relativity is discussed and astronomical
data in favor of cosmological acceleration are described. The problem of vacuum energy 
is emphasized and possible ways of its solution are indicated. The main attention is payed
to adjustment mechanism which in principle could compensate originally huge vacuum 
energy down to cosmologically acceptable value and to solve the coincidence problem of
a close magnitudes of the non-compensated remnants of vacuum energy and the energy 
density of the universe at the present time. Finally possible modifications of gravity at large
scales which could induce accelerated cosmological expansion are considered.
\end{abstract}

\maketitle

\section{Introduction \label{s-intro}}
Astronomical data accumulated during the last two or three decades unambiguously proved that 
the universe expands with acceleration, i.e. the cosmological  expansion velocity rises with time.
To appreciate the surprising features of this result let us consider a simple, though not precise, analogy.
A stone thrown vertically up from the Earth surface would either stop at some moment and return back to the 
Earth or, with sufficiently large initial speed, would fly away forever to infinity.  There is a special intermediate
case when the velocity of the stone at infinity tends to zero. In all the cases the acceleration is directed 
towards the Earth, i.e.  the velocity during "expansion" drops down. It was believed until recently that the 
cosmological expansion proceed qualitatively in the same way. The speed of the expansion decreases with 
time and either the expansion will stop and the universe will collapse back to a  hot singular state, or the 
expansion will last forever. As in the case of the stone considered above the velocity of the expansion
always drops down with time. The intermediate situation of zero velocity at infinity (i.e. expansion stops
at infinity) corresponds to spatially 
flat universe with 3D Euclidean geometry. In this old picture the spatial  geometry and the ultimate destiny 
of the universe are rigidly connected. If the universe is spatially closed (3D sphere), the expansion turns
into contraction. If the universe is open (3D hyperboloid) the expansion never stops.

It is understood now that this picture is not true. First, as is mentioned above,
the universe today expands with acceleration but most of its previous history (in terms of the cosmological 
red-shift but not time, see below) it expanded with the
normal deceleration. It is similar to a stone thrown up from the Earth surface which first moved with a decreasing
velocity but after a while it started to accelerate (as if it has a rocket accelerator), gained speed, and never 
came back. It is practically established that at the very beginning the universe also expanded with acceleration.
It is the famous inflationary stage when the initial cosmological push was operated for a short while. It is analogous
to the initial acceleration of the stone in the considered example.
After that the subsequent motion both of the stone and of the universe was simply the inertial one.

As we understand, most probably the initial inflationary push and the cosmological expansion today 
are both created by an antigravitating state of matter, though some other mechanisms are not excluded.
The antigravity at the beginning is necessary for our existence because it  was a source of expansion
which created large and suitable for life universe but it seems unnecessary now, or we do not understand its
necessity.

The paper is organized as follows. In the next section the basic cosmological equations are derived in a
simplified way and different expansion regimes are considered. In sec. 3 main cosmological parameters 
are introduced and the data in favor of the accelerated expansion are presented. In sec. 4 the problem of 
vacuum energy is described and some suggestions for its solution are discussed. In sec. 5 modified 
gravity models leading to accelerated expansion are considered.

\section{Basic equations \label{s-basic-eqn}}

The distribution of matter in the universe is assumed to be homogeneous and isotropic, 
at least at an early stage, as indicated by the isotropy of the Cosmic Microwave Background 
radiation (CMB), and even now at very large scales, larger than 100 Mpc.
Correspondingly the metric can be chosen in the Friedman-Robertson-Walker (FRW) form:
 \be
ds^2 = dt^2 - a^2(t)\,\left[f(r) dr^2 + r^2d\Omega \right],
\label{ds2}
\ee
where ${f(r)}$ describes three-dimensional (3D) space of constant curvature,
${{ f(r) = 1/(1-k r^2)}}$ with $k=+1$ corresponding to 3D hyperboloid, $k=-1$ corresponding to
3D sphere, and $k=0$ corresponding to flat 3D space.

{The evolution of scale factor ${ a(t)}$, i.e. the 
expansion law, is determined by the {Friedmann equations,} 
which follow from the general General Relativity (GR) ones: 
\be  
R_{\mu\nu} - \frac{1}{2} g_{\mu\nu} R = \frac{8 \pi}{m_{Pl}^2}\, T_{\mu\nu}
\label{R-mu-nu}
\ee
for  FRW anzats (\ref{ds2}).  The term in the r.h.s. of this equations is the energy-momentum
tensor of matter, which is the source of gravity in GR. This equation has a very important property
that its left hand side (l.h.s.) is covariantly conserved:
\be
D_\mu \left(R^\mu_\nu - \frac{1}{2} \delta^{\mu}_{\nu} R \right) \equiv 0\,,
\label {bianchi}
\ee
where $D_\mu$ is the covariant derivative in the corresponding Riemann space-time. 
This identity is the so called Bianchi identity valid in any metric theory. The conservation of the 
l.h.s. implies the conservation of the r.h.s.:
\be
D_\mu T^\mu_\nu =0.
\label {D-mu-T}
\ee
On the other hand this law of covariant conservation of the energy-momentum tensor is fulfilled due to
invariance of the theory with respect to general coordinate transformation, i.e. general covariance. In this
sense the theory is self-consistent.

The derivation of the Friedmann equations from Einstein equations (\ref{R-mu-nu}) is 
straightforward  but quite tedious. So we proceed in a simplified, non-rigorous  way which not 
only allows to avoid complicated algebra but gives an intuitive understanding of the
equations. Let us consider a small sphere in the cosmological matter with radius $a$
and a test body on the surface of this sphere. The matter outside this
sphere does not have any gravitational action on the test body, so the law of the energy 
conservation for such test body can be written as
\be
E_{kin} + U =  v^2/2- G_N\,M/a = const\,,
\label{E-conserv}
\ee
where $G_N\equiv 1/m_{Pl}^2$ is the gravitational coupling constant,
${ M= 4\pi a^3 \rho/3} $ is the total mass inside radius $a$, and 
$\rho$ is the energy density of matter inside the sphere.

Introducing the Hubble parameter $H= \dot a /a$ we rewrite this equation in the
canonical form:
\be 
H^2 \equiv \left(\frac{\dot{a}}{a}\right)^2 =
 \frac{8\pi\,\rho\, G_N }{3} - \frac{k}{a^2}
\label{H2}
\ee
From this definition and the form of metric (\ref{ds2}) follows the Hubble expansion law:
\be
v = H d \,,
\label{hubble-law}
\ee
where $v$ is the velocity of two non-interacting objects at distance $d$.
Note that objects in bound systems e.g. (stars in galaxies)) do not 
flow away from each other. The Hubble expansion law is simply motion by inertia of
previously accelerated and now non-interacting (except for gravity) objects.

One more equation follows from the energy balance in volume element $V$ of the
medium: ${ dE = -P\,dV}$, where $P$ is the pressure density and
$ { E =\rho V}$, so that  $ dE = V d\rho + 3 (da/a) V\rho $. Hence: 
\be
\dot{\rho}+3H(\rho+P)=0.
\label{dot-rho}
\ee
This equation is the condition of the covariant conservation of the energy-momentum tensor
(\ref{D-mu-T}) for a special case of FRW metric.

From eqs. (\ref{H2}) and (\ref{dot-rho}) follows the expression
for the acceleration of the test body:
\be
\frac{\ddot a}{a} = -\frac{4\pi\,G_N }{3}\,(\rho+3P)\,.
\label{ddot-a}
\ee
{Let us stress that not only energy but also pressure gravitate and induce gravitational
acceleration.} Here lays a fundamental difference between the General Relativity and the non-relativistic
Newtonian theory of gravity. 

It looks quite surprising that our non-relativistic analysis of the behavior of the test body in seemingly 
Newtonian gravitational theory have lead us to GR equation ({\ref{ddot-a}). The essential input, where GR 
is involved, is the assumption that the source of gravity is not mass but  energy. Together with the law of the
energy balance/conservation (\ref{dot-rho}) it leads to participation of pressure in creation of the 
gravitational force and this is of fundamental importance for existence of life. The energy density of matter
must be positive in any known reasonable non-pathological theory. So if only energy density gravitated, 
gravitational repulsion would be impossible and the universe would not expand to a possible-for-life state.
However, pressure may be negative and if $P < - \rho/3$, acceleration $\ddot a$ would be positive and it
would result in cosmological antigravity and induce cosmological expansion. 
Life is impossible with Newtonian gravity.

Existence of gravitational repulsion in cosmology does not mean that it may be possible to construct  an
anti-gravitating space ship. Antigravity is allowed only for infinite size objects, such as the universe as a
whole, domain walls, or cosmic strings. One can see that finite objects can create only gravitational attraction
analyzing the Schwarzschild solution, according to which the gravitational field created by a compact object
is proportional to a  positive definite integral of the energy density multiplied by a certain positive function which
includes the gravitational mass defect. The mass, which creates the  gravitational field,
\be 
M = \int dr r^2 T_0^0\,,
\label{M}
\ee
is positive for any pressure.

Another way to check that the pressure inside a compact source does not gravitate is to integrate by parts the
expression:
\be
0 = \int d^3 x \partial_j (x^l T^j_k) = \int d^3 x T_k^l \,.
\label{grav-pressure}
\ee

To complete the set of the equations describing the cosmological evolution we need to know 
the equation of motion of a test particle in the FRW space-time.  If  3D  space is flat, the variation of
the particle momentum is simply determined
by the Doppler effect. Indeed, let us consider a particle moving from point $x_1$ to $x_2$ separated 
by distance $dx=x_2-x_1$.  According to the Hubble law (\ref{hubble-law}), the relative velocity of these two points is 
{${ U= H dx}$}. Correspondingly the Doppler shift of the  particle momentum is $dp = -UE = -HE dx $ and so:
\be 
\dot p = -HE \dot x = -Hp\,.
\label{dot-p}
\ee
In other words, the momentum of a free particle drops down as the inverse scale factor. At this stage it is
convenient to introduce the cosmological redshift:
\be
z = a(t_U)/a(t) - 1\,.
\label{z}
\ee
So the solution of eq. ({\ref{dot-p}) looks as: {${ p \sim 1/a \sim 1/(z+1)}$.}

Eq. (\ref{dot-p}) is the geodesic equation in 3D flat FRW metric. The general form of the
geodesic equation is:
\be 
\frac{d V^\alpha}{ds} = - \Gamma_{\mu\nu}^\alpha V^\mu V^\nu \,,
\label{dV-ds}
\ee
where ${V^\alpha = dx^\alpha/ds}$ and $\Gamma^\alpha_{\mu\nu}$ are the Christoffel symbols in FRW metric. 
If the 3D space is not flat, there appeared some additional terms
proportional to spatial curvature, which usually are negligibly small.

Out of three equations  (\ref{H2}), (\ref{dot-rho}), and (\ref{ddot-a}) only two are independent but they contain three unknown
functions of time: $\rho, P$, and $a$. Hence for the description of the cosmological evolution one more
equation is necessary. Normally to this end the equation of state is used defining pressure density
in terms of energy density: $ P = P(\rho)$. In many practically interesting cases the linear equation of
state is sufficient:
\be
P = w \rho\,\ 
\label{eq-of-s}
\ee
where $w$ is (normally) a constant parameter, different for different forms of matter. 

Now we discuss three 
equations of state which were realized in cosmology at different periods of the universe evolution.

For nonrelativistic matter $P\ll \rho$ . Hence in a good approximation 
one can take  $w =0$. So from eq. (\ref{dot-rho}) follows
$ \dot \rho_{NR} = -3H\rho_{NR} $ and  thus:
\be
 \rho \sim 1/a^3\,.
\label{rho-NR-of-a}
\ee
The result is physically evident: the number density of the particles drops down
as volume (as it is sometimes said that the number density is conserved in comoving volume, i.e. in the
volume expanded together with the universe), while the energy of each nonrelativistic particle is equal to 
its mass and remains constant. 

The expansion law can be found from equation  (\ref{H2}) and in the particular case of $k=0$ has a very 
simple form:
\be 
a_{NR} \sim t^{2/3}.
\label{a-NR}
\ee

For relativistic matter {${ P=\rho/3}$.} This expression is intuitively evident because for relativistic particle
momentum is equal to its energy and the pressure is obtained by averaging of the momentum flux over 
three spatial directions. Correspondingly, according to eq. (\ref{dot-rho}) the energy density  evolves as
\be
{ \rho_{rel} \sim 1/a^4}\,.
\label{rho-rel-of-a}
\ee
An extra factor $1/a$ in comparison with nonrelativistic matter
appears because of the energy/momentum redshift of individual relativistic particles. 
Correspondingly in relativistic expansion regime {${ a\sim t^{1/2}}.$}

Another very interesting and rather unusual equation of state is the so called 
vacuum(-like) one, when $w=-1$, which is (approximately) realized in the universe at the present time.
Its source  can be in particular the vacuum energy-momentum tensor. Since vacuum is supposed to be
the same independently of the reference frame, its energy-momentum tensor
must have the same form in any reference frame too and thus it  must be proportional to 
the only "invariant" second rank tensor, i.e. to metric tensor, $g_{\mu\nu}$:
\be
T_{\mu\nu} =\rho_{vac}\,\, g_{\mu\nu}\,\,\,\, {\rm and}\,\,\,\,
P_{vac} = -\rho_{vac} \,.
\label{T-mu-nu-vac}
\ee
Since {${ \dot \rho = -3H(\rho+P) = 0}$}, 
vacuum energy density remains constant in the course of the cosmological expansion: 
\be
\rho_{vac} = const\,.
\label{rho-vac-of-a}
\ee 
With constant energy density the Hubble parameter is constant too and the expansion is exponential,
${ a\sim \exp (H_{vac}t)}$.

Such strange behavior of the energy density which does not decrease, while the volume rises exponentially,
leads to a very important and striking phenomenon. Namely if in some piece of the "primordial" universe, 
the equation of state would be the vacuum-like one, (\ref{T-mu-nu-vac}), this presumably microscopically 
small volume would quickly expand 
becoming larger than the present-day universe  and all the matter in our universe 
would be created from the stretched out microscopically small amount of matter in the original volume.
So all our visible universe could originate from microscopically 
small volume with negligible amount of matter {by exponential expansion with constant
${\rho}$.}

Note that if ${ k>0}$ and ${ \rho \sim 1/a^n}$ with n=3,4, the cosmological expansion will change 
to contraction. If $k<0$, 
the expansion will be eternal for any equation of state. However, if   ${\rho >k/a^2}$, as e.g. $\rho_{vac}=const$,  the 
expansion will last forever for any $ k $.

As we have seen, the set of equations (\ref{H2}), (\ref{dot-rho}), and  (\ref{ddot-a})  does 
not have stationary solutions. 
It was to great  disappointment of Einstein who wanted to have static universe despite the problem of thermal
death of the universe and the Olberts paradox.  So Friedman~\cite{friedman}, who first found the non-stationary
realistic cosmological solutions predicted the universe 
expansion, later discovered by Le Maitre~\cite{lemaitre} and Hubble~\cite{hubble}.
Thus the Hubble law would be more appropriate to name the Friedman-LeMaitre-Hubble law (FMH law). 
After the works of Robertson and Walker~\cite{RW}, who studied general properties of the metric
associated with the Friedman solution, it got the name FRW metric.

Note in conclusion to this section that if $w< -1$, the energy density rises with time and such behavior  leads
to the so called phantom singularity:
\be
a(t) \sim \left( \frac{ 1}{t_0-t} \right)^{2(|w|-1)/3}\,, \,\,\, \rho \sim   \left( \frac{ 1}{t_0-t} \right)^{2(|w|-1)} \,.
\label{sing-w}
\ee
Thus in finite time both the scale factor and the energy density would become infinitely large and 
not only galaxies, stars, and planets, but atoms and even elementary particles would be turn apart.
The existing astronomical data indicate to the possibility that $w <-1$ but it does not necessarily means that the universe will
end in singular state because $w$ may change with time and return to $w\geq -1$ before the singularity is reached.
For example $w(t)<-1$ may be realized by a scalar field with some unusual form of the kinetic term, 
but in the course of the
field evolution $(w+1)$ may become non-negative. 

The energy-momentum tensor of a scalar field with the normal kinetic term $\phi$ has the form:
\be
T_{\mu\nu} = \partial_\mu \phi \partial_\nu \phi -
\frac{1}{2}g_{\mu\nu} \left[ (\partial \phi)^2 - U(\phi)\right]
\label{T-mu-nu-of-phi}
\ee
Correspondingly for a homogeneous scalar field $\phi = \phi (t)$ the energy and pressure
density are:
\be
\rho = \frac{\dot \phi^2 + U(\phi}{2}, \,\,\,\, P = \frac{\dot \phi^2 - U(\phi)}{2}
\label{rho-P}
\ee
and the condition $|P|<\rho$ is always fulfilled, if pressure is negative.
Clearly for a slowly varying $\phi$ the vacuum-like equation of state $P \approx -\rho$ is realized. In the
limit of very fast field variation when the potential can be neglected the most rigid equation of state $P=\rho$
becomes valid. 

Note that the equation of state in the form $P = P (\rho)$ does not always exist. The relation between $P$ and $\rho$ 
may be non-local in time, though one can always define parameter $w(t) \equiv P(t)/\rho(t)$. Despite an absence of
the equation of state, the set of cosmological equation can be still complete because the necessary missing equation
is the equation of motion for $\phi$:
\be 
D^2 \phi + U'(\phi) = 0\,.
\ee

\section{Universe today \label{s-univ-today}}

The rate of the universe expansion, see eq. (\ref{hubble-law}), is determined by the present day value of the Hubble parameter:
\be
H= 100\, h\,{\rm km/sec/Mps}\,,
\label{H0}
\ee
where ${ h= 0.73 \pm 0.05}$ and ${H^{-1}= 9.8\,{\rm Gyr}/h \approx 13.4\, {\rm Gyr}.}$

The cosmological energy densities of different forms of matter, $\rho_a$ are expressed through dimensionless 
parameters:
\be 
\Omega_a = \rho_a /\rho_c,
\label{Omega-a}
\ee
where $\rho_c$ is the so called critical energy density, i.e. energy density in spatially flat universe corresponding to 
$k=0$. From equation (\ref{H2}) it follows:
\be
\rho_c = \frac{3H^2 m_{Pl}^2}{8\pi}=
1.88\cdot 10^{-29} h^2 { { \frac{g}{\rm cm^3}}}=
10.5\, h^2  {\rm {\frac{keV}{cm^3}}} \approx 10^{-47} h^2\,{\rm { GeV^4}}\,.
\ee 
It corresponds to 10 would-be protons per ${ m^3}$, but in reality the density of protons is 
much smaller; the dominant form of matter in the universe is not the usual baryonic one but something
invisible called dark matter and dark energy.

{\it Matter inventory} (for a review see e.g.  ref. \cite{matter-rev}):\\
The total cosmological energy density is quite close to the critical one:
\be 
 \Omega_{tot}  = \sum_a \rho_a /\rho_c = 1 \pm 0.02\,.
 \label{Omega-tot}
 \ee
This result is obtained from the analysis of the angular fluctuations of CMBR, in
particular, from the the position of the first acoustic peak peak and from the study of the
large scale structure (LSS) of the universe.

The usual baryonic matter contributes a minor fraction to the cosmological energy density,
$ \Omega_{B} = 0.044 \pm 0.004$. 
It is measured by comparison of the relative heights of CMBR peaks and by the analysis of big bang 
nucleosynthesis (BBN). Qualitatively one can conclude that the fraction of baryons is small because
in the baryon dominated universe the structure formation could start only rather late, after recombination at 
$z\approx 10^3$, but in this case there is too little time for density perturbations to evolve up to the
observed values, $\delta\rho/\rho \geq 1$.

Thus 95\% of matter in the universe is something unknown, dark. This dark staff can be separated into two 
quite different components:
dark matter (DM) and dark energy (DE). Dark matter is supposed to consist of some unknown objects (new
stable elementary particles, primordial black holes, topological or non-topological solitons, etc). The fraction of 
dark matter is five times larger than fraction of baryons:
{${{ \Omega_{DM} \approx 0.22\pm 0.04 }}$}. 
This number was obtained from galactic rotation curves, gravitational lensing, equilibrium
of hot gas in rich galactic clusters, cluster evolution, and baryon acoustic oscillations.
The equation of  state of dark matter is  simple non-relativistic one with $w=0$.

The remaining three fourth of the cosmological energy density consists of mysterious dark  energy
with $ \Omega_{DE} \approx 0.75$  and negative parameter of the equation of state, $ w \approx -1$.
According to eq. (\ref{ddot-a}),  such state of matter induces accelerated expansion. The cosmological
fraction of DE was found from the dimming of high-z supernovae, CMBR, LSS, and from the universe age.

We can calculate the universe age in terms of the present day values of the
cosmological parameters integrating the equation:
\be
\dot a  =\left[
 {8\pi\,\rho\, G_N\, a^2}/{3} - {k}\right]^{1/2}\,.
\ee
As a result one obtains
\be
t_U = \frac{1}{H}\,\int_0^1 \frac{dx}
{\sqrt{1-\Omega_{tot}  + {\Omega_m}/{x}
+{\Omega_r}/{x^2} + x^2\Omega_v }}\,,
\label{tU}
\ee
where $\Omega_m$ and $\Omega_r$  are the fractions of non-relativistic and relativistic matter respectively,
and $\Omega_v$ is the fraction of vacuum energy with $P=-\rho$. 

According to the ages of old stellar globular clusters and nuclear chronology the universe
age lays in the interval:
\be
t_U = 12-15 \,{\rm Gyr}\,,
\ee
which well agrees with the presented above data on $H$ and $\Omega_a$ but disagrees with decelerated
universe without dark energy.

Let us stress that different methods of determination of the cosmological parameters are absolutely 
independent. They are obtained by different types of astronomical observations and rely on different
physical phenomena in the universe. The latter excludes possible interpretation errors. So 
phenomenologically it is established without any doubts that the universe expands with acceleration and
the source of this accelerated expansion makes 75\% of the total cosmological energy density but 
it is unknown what exactly it is. There are two main competing possibilities: a new light or massless (scalar)
field or modification of gravity at cosmologically large scales. The first one includes in particular some 
not completely compensated remnant of vacuum energy carried by the compensating field. 
In the case of modified gravity one must take care not only of the dimming of the high ref-shift
supernovae but also of the large scale structure formation, CMB fluctuations, universe age, etc.

\section{Problem of vacuum energy \label{s-vac-en}}

The problem of vacuum energy is probably the most striking problem of contemporary 
fundamental physics. This is a unique example when theoretical expectations differ from
the observations by 50 -100 orders of magnitude. The beginning of the story is traces back
to almost century ago~\cite{ein-lambda} when Einstein introduced into GR equations (\ref{R-mu-nu})
an additional term proportional to metric tensor $g_{\mu\nu}$:
\be 
R_{\mu\nu} -\frac{1}{2} g_{\mu\nu}R - \Lambda\, g_{\mu\nu} = 
8\pi G_N\,T_{\mu\nu}\,. 
\ee
Coefficient $\Lambda$ must be constant to satisfy the constraints of general covariance and 
energy-momentum conservation, see eqs. (\ref{bianchi}) and (\ref{D-mu-T}). By this reason
Lambda-term is often called cosmological constant.

Evidently ${{ \Lambda}}$-term is equivalent to vacuum energy (\ref{T-mu-nu-vac}), though there are still
some statements in the literature that $\Lambda$ is a geometrical quantity, while $T_{\mu\nu}^{(vac)} $ 
has completely different nature and so they should be treated differently. However, there is no way to 
distinguish them observationally and thus they are absolutely the same from physical point of view.

As follows from quantum field theory, vacuum energy of any quantum field is infinitely
large. Indeed quantum field can be understood as a collection of quantum oscillators with all
possible frequencies each having the ground state energy equal to $\omega/2$. Integration over
all frequencies for a bosonic field gives:
\be 
\rho_{vac}^{(b)} \equiv \langle {\cal H}_b \rangle_{vac} = \int \frac{d^3 k}{(2\pi)^3}\,
\frac{\omega_k}{2}
\langle a^\dagger_k a_k + b_k b^\dagger_k \rangle_{vac} \,
 =\int \frac{d^3k}{(2\pi)^3}\,\omega_k = {\infty^4}\,.
\label{h-b}
\ee
Here $\omega_k = \sqrt{k^2 + m^2}$ and $m$ is the mass of the field.

On the other hand, fermionic vacuum fluctuations also have infinitely large vacuum energy but of the opposite sign:
\be 
\rho_{vac}^{(f)} \equiv \langle {\cal H}_f \rangle_{vac} = \int \frac{d^3 k}{(2\pi)^3}\,
\frac{\omega_k}{2}
\langle a^\dagger_k a_k - b_k b^\dagger_k \rangle_{vac} \,
= \int \frac{d^3k}{(2\pi)^3}\,\omega_k = -\infty^4\,.
\label{h-f}
\ee

It was noticed by Zeldovich~\cite{zeld-susy} that in the world with an equal number of bosonic and fermionic
species having equal masses, at least pairwise, energy of vacuum fluctuations vanishes. Later it was suggested
that  there may indeed exist symmetry between bosons and fermions called supersymmetry (SUSY)~\cite{golfand},
which demands an equal number of bosonic and fermionic degrees of freedom. If so, then the quartically divergent
contribution to vacuum energy vanishes. Clearly supersymmetry is not exact because there are no supersymmetric 
partners of the observed particles with the same masses. Presumably they are much heavier, at least by 
$m_{SUSY} \sim 10^3 - 10^4 $ GeV. In this case the sum of bosonic (\ref{h-b}) and fermionic (\ref{h-f}) contributions
would be "only" quadratically infinite, $\rho_{vac}^{(b)} + \rho_{vac}^{(f)} \sim m_{SUSY}^2 \times \infty^2 $. With a particular
mechanism of SUSY breaking the inifinities may be cancelled but the net result must be non-zero and of the order of
$m_{SUSY}^4$, which is larger than the observed value of   $\rho_{vac}$  by  at least 56 orders of magnitude. In the
version of supersymmetry which includes gravity (supergravity) the net result for vacuum energy may vanish in
the broken case, but at the expense of a fantastic fine-tuning. The natural 
value of $\rho_{vac}$ in supergravity is about
$m_{Pl}^4 \sim 10^{76}$ GeV$^4$, so the fine-tuning of the parameters must be achieved with 
the unbelievable precision $10^{-123}$.

Let us note also that vacuum (or vacuum-like) energy undergoes several huge jumps in the course of
the cosmological evolution. At the very beginning during inflationary epoch the vacuum-like energy of the
inflaton field could be as large as $10^{56}$ GeV$^4$. Though it was not real vacuum energy but still the
inflaton was anti-gravitating similar to the present today dark energy. Later in the process of 
the cosmological cooling down
a sequence of phase transitions occurred in the primeval plasma. If at the end of inflation the plasma was 
heated up to the temperatures above the grand unification scale, $T\sim m_{GUT} \sim 10^{15}$ GeV, the GUT symmetry
would be unbroken and broke downs later at smaller $T$. The phase transition from symmetric to broken
symmetry state is accompanied by the change of the ground state energy-momentum tensor 
approximately equal to $\Delta T_{\mu\nu} \approx m_{GUT}^4 g_{\mu\nu}$. 
Similar phase transitions took place at the electroweak and QCD scales with change of vacuum energy
$10^{8}$ GeV$^4$ and $10^{-2}$ GeV$^4$ respectively. There might also be phase transitions related
to supersymmetry breaking.

These simple estimates show that vacuum energy is naturally expected to be either infinite or almost infinite.
The attitude of the community, at least of that part of the community which appreciated this problem was quite 
strange, namely it was implicitly taken that $\infty = 0$. Such approach reminds well known quotation from Feynman 
about radiative corrections in quantum electrodynamics: ``Corrections are infinite but small''. 

The attitude of the astronomers and cosmologists to Lambda term was (and still is) strongly polarized. The dominant 
part of the establishment believed that Lambda is identically zero, while some other thought that introduction of Lambda
term was a very important generalization of GR~\cite{lambda-OK}. 
Among the last was  Lemaitre who said that it was the greatest discovery, 
worth alone to make Einstein's name famous. 
A strong antagonist of Lambda was Gamow, who wrote in his autobiography
book "My world line"~\cite{gamow} that ${\lambda}$ raises its nasty head again"
after indication at the beginning of the 60s that quasar are  accumulated near $z=2$, which later found
"Lambda-less" explanation.

Now the gravity of the vacuum energy problem placed it into the central position in fundamental physics. 
Probably the  most serious argument in favor that something mysterious happens in vacuum comes  from
quantum chromodynamics (QCD). According to this well established theory which  beautifully passed all
experimental tests, $u$ and $d$ quarks are very light. Their masses are about  5 MeV.
Proton is known to be a bound state of these three quarks, $p= (uud)$.
So the proton mass should be 15 MeV minus binding energy, instead of 938 MeV.
The solution of the problem suggested by QCD is that
{vacuum is not empty} but filled with quark~\cite{q-cond} and gluon~\cite{g-cond} condensates:
\be 
\langle \bar q q {\rangle} &{ \neq}&{ 0}\,, \nonumber \\
{ \langle}{ G_{\mu\nu} G^{\mu\nu} }{ \rangle} &{ \neq}&{ 0}\,,
\ee\
having negative vacuum  energy:
\be 
\rho_{vac}^{(QCD)} \approx\,(-0.01 \,{\rm GeV}^4) \approx (- 10^{45} \rho_c)\,.
\label{rho-vac-QCD}
\ee
Vacuum condensate is destroyed around quarks and the
proton mass becomes
\be
m_p = 2m_u + m_d - \rho_{vac}^{(QCD)} l_p^3 \sim 1 \,{\rm GeV}\,,
\ee
where $l_p \sim (0.1 {\rm GeV}^{-1} )$ is the proton size.

The value of the vacuum energy of the quark and gluon condensates (\ref{rho-vac-QCD}) is practically
established by experiment. To adjust the total vacuum energy down to the observed magnitude, 
$\sim 10^{-47} $ GeV$^4$, there must exist another contribution to vacuum energy of the opposite
sign and equal to the QCD one with precision of one part to $10^{45}$. This new field cannot have
any noticeable interactions with quarks and gluons, otherwise it would be observed in direct experiment,
but still it must have very same vacuum energy. This is one of the greatest mysteries of Nature.
 
At this place it is proper to make an intermediate summary.\\
1. There are known and fantastically huge contributions to $\rho_{vac}$ but a mechanism of their
compensation down to (almost) zero remains mysterious.\\
2. The observed today vacuum energy is very close the energy density of the cosmological matter, 
$\rho_{vac} \sim 3 (\rho_B+\rho_{DM} )$, 
though they evolve in a different way during cosmological history, i.e. 
$\rho_{matter} \sim 1/t^2$ and $\rho_{vac} = const$. 
Why their near equality happened to occure at the present epoch? \\
3. What is the nature of anti-gravitating dark energy? It seems to have the equation of state consistent with 
${{ w=-1}}$. Is it simply vacuum energy or something more interesting?

Mostly only problems 2 and 3 are addressed theoretically (phenomenologically) either by
(infrared) modification of gravity or by a
new (scalar) field (quintessence) leading to the accelerated expansion.
However  evidently all three problems are strongly coupled and can be solved
only after adjustment of "infinitely" large  ${\rho_{vac}}$ down to tiny ${\rho_c}$ is understood.

\subsection{Some suggestions for the solution of the vacuum energy problem \label{ss-solutions}} 

There are several suggestions in the literature for possible resolution of the vacuum energy 
problem, for a review see ref.~\cite{lambda-rev}.
Below we briefly discuss some of them and dwell in some detail on the dynamical adjustment
mechanism. The short and incomplete list of different option looks as follows:\\
{\it 1. Subtraction constant.} If we encounter exact vacuum energy problem, i.e. $w = -1$, a trivial
resolution of the difficulty could be a very precise choice of the zero level of the energy density or,
in other words, the choice of the subtraction constant whose  value must be taken such that it would
very accurately compensate all the contributions to vacuum energy from vacuum fluctuations, phase 
transitions, etc, being taken with opposite sign to them, but with small "imperfection" exactly equal to the
observed value of $\rho_{vac}$. It is impossible to exclude such "solution" but it is extremely
unnatural. 

If ${{ w\neq -1}}$, dark energy is not constant but evolves during cosmological history and
simple subtraction cannot eliminate it.\\ 
{\it 2. Anthropic principle}~\cite{anthrop}. 
In some sense it is similar to the solution with help of the subtraction constant. 
Unnaturalness of the latter is lifted by the assumption that there is a huge number of different 
universes with all possible sets of the subtraction constants. The conditions suitable for life 
exist only in universes with rather narrow span of values of $\rho_{vac}$. Large and positive $\rho_{vac}$ 
would produce too fast cosmological expansion, so the structure formation would be strongly inhibited. The opposite
case of negative and large by the absolute value $\rho_{vac}$ would force the universe to collapse back to
hot singularity before it becomes suitable for life. An existence of many universes with different physical
laws and possibly with different subtraction constants in connection with the vcuum energy problem was
suggested in ref.~\cite{sakharov-lambda}. Multivacuum states can be realized in brane landscape theories~\cite{landscape} 
with about $ 10^{1000}$ different vacuum states. Large number of vacuum states naturally appear in inflationary
scenarios~\cite{infl-multi}, especially in chaotic inflation~\cite{chaotic}. Realistic way of compactification in 
multidimensional brane theories leading to huge number of vacuum states was suggested in ref.~\cite{compact}.

For many people (including the author) the anthropic solution looks unsatisfactory because the theory is impossible
to falsify. It reminds the situation with the problems of the Friedmann cosmology prior to the suggestion of  the  
inflationary paradigm. Before inflation the anthropic principle was believed to be the only feasible way to understand
how or why our universe was created. Inflation not only resolved all the problems in a simple and natural way
but also predicted some observational effects, in particular, the shape 
of the perturbation spectrum~\cite{mukh-spectr}, which is confirmed by the data,
and cosmic background of relic gravitational waves~\cite{starob-gw}, which may be a final proof of
inflation.\\

{\it 3. Dynamical adjustment.} The idea is similar to axionic solution of the strong CP problem. A new light field
is introduced whose back-reaction cancels out the impact of the source which created this field. This principle
is known in chemistry and physics from the XIX century and has the name "the Le Chatelier principle". 

The original~\cite{ad-get-rid} and many subsequent suggestions~\cite{lambda-rev}
were based on a scalar field but higher spin fields: vector~\cite{ad-vector} or tensor~\cite{ad-tensor}, 
are also possible. %~\cite{ad-high-spin, klinkhammer}. 
The main idea is that a new field ${\Phi}$ is coupled to gravity in such a way that vacuum 
energy through its gravitational action leads to formation of the condensate of ${\Phi}$.
The energy density, $ \rho_\Phi$, compensates the original vacuum energy. If $\rho_{vac}> 0$,
then $\rho_\Phi$ should not be positive definite but the related instability should not be catastrophic 
i.e. the energy of $\Phi$ does not quickly drop down to negative infinity.

Independently on a concrete realization of the idea the generic predictions are similar and quite 
attractive. First, when the contribution of $\Phi$ on the evolution of the scale factor 
is taken into account, the exponential cosmological
expansion turns into the power law one. Second, the compensation of the vacuum energy is not 
complete but it is compensated only down to the terms of the order of ${\rho_c (t)}$. 
Third, the non-compensated energy may have quite an unusual equation of state.
Such mechanism not only solves the problem of compensation of the original vacuum energy, 
which is estimated to have a typical value of the particle physics scale, down
to a cosmologically small value and also explains the so called coincidence problem, i.e. the
close proximity of the observed $\rho_{vac}$ to the time dependent $\rho_c (t)$. 
This is exactly what is observed. In this sense the dark energy was predicted in 1982~\cite{ad-get-rid}.
Unfortunately,  despite the numerous attempts no realistic model was found starting from 1982 till now.

Probably the first person who suggested that vacuum energy may be of the order of the time dependent
cosmological energy density was Matvey Bronshtein~\cite{mb}. However, taken literally, the models  with 
${{ \Lambda = \Lambda (t)}}$ are not innocent because the transversality of the Einstein equations
demands $\Lambda = const$, see eq.~(\ref{rho-vac-of-a}). To achieve such transversality, or what is the same,
the conservation of $T_{\mu\nu}$ (\ref{D-mu-T}), some new light or massless fields  are necessary.
Another possibility demands  serious modifications of the theory, e.g. non-metric theories.
This was a reason why the Bronstein suggestion was 
strongly criticized by Landau.
However, an approximate relation ${P \approx -\rho}$ can be achieved
with light scalar field which might have ${\rho = \rho(t)}$, see eq.~(\ref{rho-P}). 

%dsp

The first model of dynamical adjustment suggested in ref.~\cite{ad-get-rid} was based on 
non-minimally coupled scalar field satisfying the equation of motion:
\be
\ddot \phi + 3H \dot \phi +U'(\phi, R) = 0\,.
\ee
The non-minimal coupling to gravity was taken in the simplest form ${ U = \xi R \phi^2/2}$.
 It is easy to see that the solutions of this equation are unstable if ${\xi R<0}$ because
 in De Sitter background this term behaves as tachyon with negative effective mass 
squared,  $m_{eff}^2 <0$. One can find that
 in the initially de Sitter state with the exponentially rising scale factor, $a(t) \sim \exp (H_v t)$,
 $\phi$ also rises exponentially but when the back reaction of the
 rising $\phi$ on the cosmological expansion is taken into account,  the exponential rise turns into the power law one.
Asymptotically:
\be
\phi \sim t\,, \,\,\,\, a(t) \sim t^\beta\,,
\ee
where $\beta$ is a constant. In other words the initial De Sitter space-time becomes the Friedmann one,
despite non-zero vacuum energy. Still the vacuum energy is not compensated because the energy-momentum
tensor of $\phi$ is not proportional to the metric tensor
\be
T_{\mu\nu} (\phi) \neq \tilde\Lambda g_{\mu\nu}
\ee
but the change of the regime is achieved due to the  weakening of
the gravitational coupling:
\be 
G_N \sim 1/t^2 .
\ee

If such rise of ${ M_{Pl}}$ took place in the early universe and was stabilized at some later but still an early stage,
this mechanism might explain hierarchy between electro-weak and gravitational mass scales~\cite{notari}.

A no-go theorem was formulated by S. Weinberg in ref.~\cite{lambda-rev} which states that scalar field cannot 
naturally solve the vacuum energy problem because one has to impose two independent conditions on the
adjustment of the potential: $U(\phi_0)  = \rho_{vac} $ and the vanishing of the derivative of the potential 
$U'(\phi_0) =0$ at the same point $\phi=\phi_0$. However, this theorem can be circumvented by more exotic
coupling to curvature or by higher spin fields.

A model with compensating vector field, ${V_\mu}$, was proposed in ref.~\cite{ad-vector} where the Lagrangian
was taken in the form:
\be
{ L_1} = \eta \left[ F^{\mu\nu} F_{\mu\nu} /4 + (V^\mu_{;\mu})^2\right]
+\xi R m^2 \ln \left( 1 +\frac{V^2}{m^2}\right)\,.
\ee
In this theory the time component of the field, $V_t$ is unstable in the De Sitter background and the solution
behaves as:
\be
V_t \sim t + c/t\,.
\ee
The dominant part of the energy-momentum tensor of this solution is proportional to the metric tensor which
compensates the original vacuum energy:
\be
T_{\mu\nu} (V_t) = -\rho_{vac} g_{\mu\nu} + \delta T_{\mu\nu}\,,
\label{T-mu-nu-vector}
\ee
where $\delta T_{\mu\nu} $ tends to zero at large time.

In this model the gravitational coupling changes with time only logarithmically which may possibly agree with the
observed bounds on allowed time variation of $G_N$, but the cosmological expansion rate is not connected with 
the matter content of the universe as is dictated by eq. (\ref{H2}).

An interesting  possibility opens a massless second rank tensor field $ S_{\mu\nu} $~\cite{ad-tensor} with
only kinetic term in the action:
\be
{\cal L}_2 = 
\eta_1 S_{\alpha\beta;\gamma} S^{\alpha\gamma;\beta}
+\eta_2 S^\alpha_{\beta;\alpha} S^{\gamma\beta}_{\> \>;\gamma}
+\eta_3 S^\alpha_{\alpha;\beta} S_\gamma^{\gamma;\beta}\,.
\label{deltal}
\ee
So it is a free field having minimal gravitational interaction.

The components of this field in FRW metric satisfy the following equation of motion
\be
 (\partial_t^2 + 3H\partial_t - {6H^2}) S_{tt} -2H^2 s_{jj} = 0\,,
 \label{dt2tt}
\ee
\be
 (\partial_t^2 + 3H\partial_t -{6H^2}) s_{tj} = 0\,,
\label{dt2tj}
\ee
\be
 (\partial_t^2 + 3H\partial_t -{2H^2}) s_{ij} -2H^2 \delta_{ij} S_{tt} = 0\,,
\label{dt2ij}
\ee
where  ${ s_{tj} = S_{tj}/a(t)}$ and ${ s_{ij} = S_{ij}/a^2(t)}$.
It is easy to see that time-time component, ${ S_{tt}}$, and isotropic
part of space-space component, ${ S_{ij}\sim \delta_{ij}}$, are unstable and rise with time in De Sitter background.

Though it is not immediately evident that gravitational coupling to matter in this model changes with time, it was
argued in ref.~\cite{rub-G-of-t} that it rises  as power of time.

At this stage the following comment is in order. 
Normally in a higher spin field theory the conditions are imposed eliminating lower spin components, as e.g. in vector field
theory the 3-dimensional scalar part (longitudinal component) is eliminated; in second rank tensor field the scalar and vector
parts are eliminated. Here we have the opposite situation when the scalar component of tensor field is physical and probably
an elimination of higher spin components is necessary. It is unclear if such a theory can be consistently formulated.
An example of such theory when gauge scalar field is described by ${ t}$-component of vector 
${{ V_\mu}}$ is presented in paper "Photon and Notoph" by Ogievetsky and Polubarinov~\cite{notoph}.

A model based on scalar field with ``crazy'' coupling to gravity was suggested in paper~\cite{muk-rand} and studied in some
detail in ref.~\cite{ad-kawa-2003}. It is based on the action:
\be
  A= \int d^4 x\sqrt{g} \left[ -\frac{1}{2} (R+2\Lambda)
  + F_1(R)\right.   + {{\left.
 { \frac{D_\mu \phi D^\mu \phi} { 2\,R^2}}
- U(\phi, R) \right] }}
\nonumber
\ee
where the system of units with $m_{Pl}^2/8\pi =1$ is used.

The equation of motion for ${\Phi}$ has the form:
\be
  D_\mu\left[ D^\mu \phi\,\left(\frac{1}{R}\right)^2\right]  + U'(\phi) = 0.
\label{D-phi}
\ee
It is essential that the equation has large coefficient in front of the highest derivative term, while
quite many modified theories have small coefficient in front of the highest derivative and the 
latter leads to serious problems.

It is easy to check that solutions of equation (\ref{D-phi}) asymptotically tends to the state in which
\be
R\sim \rho_{vac} + U(\phi) = 0\,.
\ee
In other words the vacuum energy is eliminated, avoiding the Weinberg no-go theorem.
The solution has some other nice  
(``almost realistic'') features, for example ${ H=1/2t}$ but this result
does not depend upon the cosmological matter content and is
unstable with respect to small fluctuations. To see that one needs to explore the gravitational 
equation of motion, in particular the one for the scalar curvature $R$,  which for a particular choice
${{ F_1 = C_1 R^2}}$ has the form:
\be 
  - R + 3 \left( \frac{1}{R}\right)^2 \left( D_\alpha \phi \right)^2
  - 4\left[ U(\phi) +\rho_{vac}\right]
- 6 D^2  \left[ 2C_1 R - \left(\frac{1}{R}\right)^2
  \frac{\left( D_\alpha \phi\right)^2}{R}\right] =
  T^{\mu}_{\mu}.
  \label{trace}
\ee

A desperate attempt to improve the model with non-analytic dependence on $R$:
\be 
  \frac{(D \phi)^2}{R^2} \rar - \frac{(D \phi)^2}{R\,|R|}
\ee
was unsuccessful, though some interesting solutions became stable. 

More general action with a scalar field: 
\be
A = \int d^4 x \sqrt{-g}
\left[-\frac{m_{Pl}^2}{16\pi}(R+2\Lambda )
%+F_1 (R) 
+ F_2 (\phi, R) D_\mu \phi D^\mu \phi
+ F_3 (\phi, R) D_\mu \phi D^\mu R - U(\phi, R) \right]
\ee
has not yet been explored.
Moreover a dependence on ${R_{\mu\nu}}$ and ${R_{\mu\nu\alpha\beta}}$ can be also included,
though such terms could lead to ghosts and tachyons in gravitational theory.
Anyhow,  it is difficult to construct a sensible theory without a guiding principle. 

Recently in a series of papers a model with two vector fields has been studied~\cite{klink}, which
according to the author's statement is more successful in the solution of the vacuum energy problem
then the previous ones.

Still, we do not have a realistic and consistent with all observations theory explaining compensation 
of the vacuum energy down to the observed small value. As a poor man substitution for that,
phenomenological approaches are developed. There are two main roads to follow: either to introduce into the
theory a new field whose equation of state has $w<-1/3$ or to modify gravity at cosmologically large scales
in such a way that it leads to accelerated expansion notwithstanding matter.

\section{Modified gravity}

The usual Einstein-Hilbert  action is linear in the curvature scalar $R$. This is the reason why the GR 
equations (\ref{R-mu-nu}) contain, as it is usual in other field theories, only second derivatives of metric despite the
fact that the action also contains second derivatives. Normally equation of motion have one order higher
derivatives than the corresponding action, since the kinetic term in the latter is 
a function of the bilinear combination the first order field
derivatives. Radiative corrections to the GR action lead to generation of higher powers of $R$ in the
effective action, $R^n$, as well as some invariant combinations of the curvature tensors, e.g. 
powers of $R_{\mu\nu} R^{\mu\nu}$. This result is true in the limit of small curvature, $R \ll m$,
where $m$ is the particle mass  in the matter loops in curved space-time. In the opposite limit
the effective action can be expanded in inverse powers of $R$. For details one may address 
book~\cite{birrel}. 

If the action differs from a simple linear GR form, the equation of motion would  be higher than the second
order one. Such equations should contain some pathological features as existence of tachyonic solutions
or ghosts. However, the theories whose action depends only on a function of the curvature scalar, $F(R)$,
are free of such pathologies because, as is known, they are equivalent to an addition of a scalar degree of 
freedom to the usual GR with the scalar field satisfying normal second order field equation. That's why
modification of gravity at large distances is mostly confined to  ${ F(R) }$  theories:
\be 
S = \frac{m_{Pl}^2}{16\pi} \int d^4 x \sqrt{-g} [R+  F(R)]+S_m\,.
\label{grav-mdf}
\ee

In ref. ~\cite{tkachev} a model with power law corrections to the Einstein action of the kind
$F(R) = c_1 R^2 + R^3/m_3^2 $ was considered. With $m_3$ of the order of the neutrino mass,
$m_3 \sim m_\nu$, such model  describes dark energy with the energy density of the order of $m_\nu^4$,
which is quite close to the observed value.

Soon after discovery of the accelerated expansion $F(R)$ theories became quite popular.
The pioneering works in this direction was done in ref.~\cite{capo}, which was closely followed by 
ref.~\cite{carrol}. In these works the singular in $R$ action:
\be
F(R) = - {\mu^4}/{R} \,,
\label{F-of-R}
\ee
has been explored with constant parameter $\mu$ chosen as
${\mu^2 \sim R_c \sim 1/t_U^2}$ to  describe the observed cosmological acceleration.

The corresponding equation of motion reads:
\be
\left(\frac{1}{\mu^4}+\frac{1}{R^2}\right)R_{\alpha\beta} -
\frac{R}{2}\left(\frac{1}{ \mu^4}-\frac{1}{R^2}\right) g_{\alpha\beta}  - 
D_{(\alpha} D_{\beta)}\left(\frac{1}{R^2}\right)
{+ g_{\alpha\beta} D_{\nu} D^{\nu}\left(\frac{1}{R^2}\right) }
{{ = \frac{8\pi\,T_{\alpha\beta}}{m_{Pl}^2 \,\mu^4} \,.
}}
\ee
Taking trace over $\mu$ and $\nu$ of this equation we obtain:
\be 
D^2 R - 3 \,\frac{(D_\alpha R)\,(D^\alpha R)}{R} = 
{{ \frac{R^2}{2}  - \frac{R^4}{6\mu^4}  - \frac {T\,R^3 }{6\mu^4}\,.
}}
\label{trace-F}
\ee
 Here ${T=8\pi T_\nu^\nu/m_{Pl}^2>0}$.
This equation has an evident solution in absence of matter $ R^2 = 3\mu^4$ which describes 
the accelerated De Sitter universe with a constant curvature scalar. 

So far, so good but 
the small coefficient, {${ \mu^4}$,} in front of the highest derivative or, what is the same, 
the large coefficient, $1/\mu^4$, in front of the non-derivative terms 
in the presented above form of the equation leads to
a strong instability in  presence of matter~\cite{ad-mk-instab}.
Indeed, let us apply equation (\ref{trace-F}) for perturbative calculations of the gravitational field of a
celestial body.

Let us look for the solution perturbatively with $R = R_0 + R_1$, where $R_0$ is the 
usual solution in the non-modified General Relativity (GR), ${ R_0 = -T }$. Since the gravitational
field is weak, the flat background metric is assumed. 
 
In vacuum,  outside the matter source  ${R}$ exponentially 
tends to zero, i.e to the GR  value, if ${\mu^4 >0}$. 
So these  modified gravity theories {agree with the
Newtonian limit of the standard gravity for sufficiently large ${\mu}$.}

Now let us consider the internal solution with time dependent matter
density. The first order correction to the GR curvature, ${R_1}$
satisfies the equation:
\be 
\ddot R_1 -\Delta R_1 - \frac{6\dot T}{T}\, \dot R_1 
+\frac{6\partial_j T}{T}\, \partial_j R_1  
{ {  +R_1 \left[ T + 3\,\frac{ (\partial_\alpha T)^2}{T^2}  
 {-\frac{T^3}{6\mu^4} }\right] } }
 = \Delta T + \frac{T^2}{2} - \frac{3 (\partial_\alpha T)^2}{T}\,,
%\,\partial^\alpha T}{T}\, .
\label{ddotr1}
\ee
where $(\partial_\alpha T)^2 = \dot T^2 - (\partial_j T)^2 $.

The last term in the square brackets in the r.h.s. leads to exponential instability of small 
fluctuations as well as to instability of gravitational field created by a regularly varying with
time mass density of the considered  body. The characteristic time of instability is:
\be
\tau=\frac{\sqrt{6}\mu^2} {T^{3/2} }
\sim 10^{-26} {\rm sec}  
\left(\frac{\rho_m}{{\rm g/ cm}^{3} }\right)^{-3/2}\, ,
\label{t-eff}
\ee
where ${\rho_m}$ is the mass density of the  body and
{$ {\mu^{-1} \sim t_u \approx 3\cdot 10^{17} \,{sec} }$. }
This is the dominant term in the equation, since 
$ T \sim (10^{3} {\rm sec})^{-2} \left({\rho_m}/{{\rm g\ cm}^{-3}}\right) $
and hence the ratio $ T^3/\mu^4 $ is huge.

Usually spatial inhomogeneities prevent or inhibit instability, as e.g. happens in the case 
of the Jeans instability. However in our case the term $\Delta R_1$ is by far smaller than
the "unstable" term. The instability would be damped only at the scales of the order 
of or below the  Compton wave length of proton.

To avoid the problem of such instability a modification of the modified gravity  has been
suggested. We will consider here some class of the models discussed in refs.~\cite{mdf-mdf}.
Some other forms of gravity modification are reviewed in ref.~\cite{odin-rev}.
The different actions suggested in works~\cite{mdf-mdf} have the form:
\begin{eqnarray}\label{eq:cr1}
F_{\rm HS}(R) &=&  - {R_{\rm vac} \over 2} {c \left({R \over R_{\rm vac}}\right)^{2n} \over 1+
c\left({R \over R_{\rm vac}}\right)^{2n}}\,,  \label{eq:cr201} \\
F_{\rm AB}(R) &=&  {\epsilon \over 2}\,
{\rm log} \left[ {\cosh\left({R \over \epsilon}-b\right) \over \cosh b} \right]  
 - \frac{R}{2}\,, \\
%\end{eqnarray}
%\be 
F(R)_S &=& \lambda R_0 \,\left[ {\left(1+ \frac{R^2}{R_0^2}\right)^{-n}} - 1 \right]\,.
\label{F-AAS}
\ee
Despite different forms these actions result in quite similar consequences. Below we 
essentially follow the analysis made in ref.~\cite{appl-bat-star}.

Introducing notation  ${f = R +F(R)}$ we can write the
field equations in the form:
\begin{equation}
f' R_{\mu}^{\nu} - \frac{f}{2}\delta_{\mu}^{\nu}+
(\delta_{\mu}^{\nu} D^2- D_{\mu} D^{\nu})f'= \frac{T_{\mu}^{\nu}}{M_{Pl}^{2}}\,, 
\label{FReq}
\end{equation}
and correspondingly their trace is:
\begin{equation} 
3 D^2 f'(R) + Rf'(R) - 2f(R)= M_{\rm Pl}^{-2}T^\mu_\mu\,.
\label{trace-eq}
\end{equation}

The condition of accelerated expansion in absence of matter is that the equation 
\be 
R f'(R) - 2 f(R) = 0 
\ee
has the solution  ${R=R_1>0}$, where $R_1$ is (approximately) constant. 

The following necessary conditions to avoid pathologies are to be satisfied:\\
{1.  Future stability of cosmological solutions:}
\begin{equation}
F'(R_1)/F''(R_1)>R_1\,.%\label{dSstab}
\end{equation}
{2. Classical and quantum stability (gravitational attraction  and absence of ghosts):}
\begin{equation}
F'(R)>0,~F''(R)>0\,. %\label{stabil}
\end{equation}
{3. Absence of matter instability:}
\begin{equation}
F'(R)>0,~F''(R)>0\,. 
\end{equation}

Despite considerable improvement these double modified versions still have some serious problems.
First of all they possess the so called past singularity: {in cosmological background with decreasing energy density
the system must evolve from a singular state} {with an infinite ${R}$.} In other
words, if we travel backward in time from a normal cosmological state, we come to a singular state with 
infinite curvature, while the energy density remains finite.

Moreover in the systems with rising mass/energy density the system evolves to a
singularity in the future~\cite{frolov,ea-ad}. Infinite value of $R$ would be reached in 
finite (short) time. Following ref.~\cite{ea-ad} let us consider version (\ref{F-AAS}) in the limit
${R \gg R_0} $, when one can approximately take:
\be 
F(R) \approx -\lambda R_0 \left[ 1 -\left(\frac{R_0}{R}\right)^{2n} \right] \,.
\label{F-large-R}
\ee
We analyze the evolution of ${R}$  in  a
massive object with time varying density, ${\rho \gg \rho_{cosm}}$.

%dsp

Gravitational field of such objects is supposed to be weak, so
the background metric is approximately flat and  
covariant derivatives can be replaced by the  flat ones. Hence equation (\ref{trace-eq}) 
takes the form:
\be
(\partial^2_t - \Delta) R -(2n+2) \frac{\dot R^2 - (\nabla R)^2}{R} +
 {{\frac{R^2}{3n(2n+1)} \left[\frac{R^{2n}}{R_0^{2n}} -(n+1) \right] }} \nonumber \\
 -\frac{R^{2n+2}}{6n(2n+1)\lambda R_0^{2n+1}} (R + T)=0\,.  
\label{eq-for-R}
\ee

The equation is very much simplified if we choose another unknown function:
${w \equiv F' = - 2n\lambda \left({R_0}/{R}\right)^{2n+1} }$ which satisfies:
\be
(\partial^2_t - \Delta) w  + U'(w) = 0\,.
\label{eq-for-w}
\ee
Here potential ${U(w)}$ is equal to:
\be 
U(w) = \frac{1}{3}\left( T - 2\lambda R_0\right) w + 
{{\frac{R_0}{3} \left[ \frac{q^\nu}{2n\nu} w^{2n\nu}+ \left(q^\nu
    +\frac{2\lambda}{q^{2n\nu} } \right) \,\frac{w^{1+2n\nu}}{1+2n\nu}\right]\,,
}}
\label{U-of-w}
\ee
{where ${\nu = 1/(2n+1)}$ and ${q= 2n\lambda}$.} 

Notice that the infinite ${R}$ singularity corresponds to {${w =0}$}. 

 {If only the dominant terms are retained 
and if the space derivatives are neglected, equation (\ref{eq-for-w}) simplifies to:}
\be 
\ddot w + T/3 - \frac{q^\nu (-R_0)}{3w^\nu}=0\,.
\label{eq-w-simple}
\ee 
Potential { {$U$}} would depend upon time,
if the mass density of the object changes with time. We parametrize it as:
\be 
T=T(t) =T_0 (1 +\kappa \tau)\, ,
\ee
where ${\tau}$ is dimensionless time introduced below.

With dimensionless quantities  
${t = \gamma \tau}$ and ${w = \beta z}$,
%}\nonumber
%\label{dimless}
%\ee
where
\be
\gamma^2 = \frac{3q}{(-R_0)} \left(-\frac{R_0}{T_0}\right)^{2(n+1)}\,,\nonumber\\
\beta = \gamma^2T_0/3 = q \left(-\frac{R_0}{T_0}\right)^{2n+1}
\label{gamma-beta}
\ee
the equation  further simplifies:
\be
z'' - z^{-\nu} + (1+\kappa \tau) = 0\,.
\label{eq-for-z}
\ee
One can easily see that the minimum of this time dependent potential moves with time closer to $z=0$
and becomes more and more shallow. So it is practically evident that independently on the initial
condition in a system with rising mass density, $z$ reaches zero in finite time and correspondingly
$ R = \infty $. Numerical calculations of ref.~\cite{ea-ad} show exactly such a picture.
To cure this ill-behavior one may add into the action quadratic in curvature term 
$R^2/6 m^2$~\cite{aas-R2}, which does not allow $R$ to reach infinity but the system is
stabilized at quite large values of $R$.
In the homogeneous case and in the limit
of big ratio $R/R_0$ the equation of motion with $R^2$ addition is modified as
\be
\left[ 1-\frac{R^{2n+2}}{6\lambda n(2n+1) R_0^{2n+1} m^2 }\right]\,\ddot R 
- (2n+2) \,\frac{\dot R^2}{R} -
\frac{R^{2n+2} (R+T)}{6\lambda n (2n+1) R_0^{2n+1}} = 0 \,.
\label{eq-for-R-mdf}
\ee
An important effect which in not taken into account in this equation and which
also inhibits unbounded rise of $R$ is the particle production by oscillating 
curvature $R$. The technique for calculations of particles production applicable to the case 
of modified gravity (\ref{eq:cr1}-\ref{F-AAS}) was worked out in ref.~\cite{ea-ad-lr} 
for the case of $R^2$ gravity in cosmological situation (in the early universe) 
where the classical results~\cite{pp-cosm} for particle production were reproduced. 
The developed method  is applied to particle production in the modern astronomical systems
 which under certain conditions may be observable sources of cosmic rays~\cite{ea-ad-lr-2}.
 
To conclude, there are quite many phenomenological models but no understanding  of 
the cosmological acceleration and of the vacuum energy compensation mechanisms are found 
at a deeper level. A very important for future development would be an accurate
measurement of $w$. For vacuum energy $w$ must be strictly equal to (-1). The modern data 
indicate that $w$ is close to this value but some deviation, which could be crucial, is possible.

{\bf Acknowledgement.}
This work was supported by the Grant of the Government  of the Russian Federation
No. 11.G34.31.0047.

% ----------------------------------------------------------------
% The end of the bibliography
% ----------------------------------------------------------------

%

% \end{figure}

\end{document}